# Cross-Corpus Validation of Speech Emotion Recognition in Urdu using Domain-Knowledge Acoustic Features

Unzela Talpur [1], Zafi Sherhan Syed [2], Muhammad Shehram Shah Syed, Abbas Shah Syed

{ [1] Unzelatalpur@gmail.com, [2] zafisherhan.shah@faculty.muet.edu.pk}

*Abstract:* Speech Emotion Recognition (SER) is a key affective computing technology that enables emotionally intelligent artificial intelligence. While SER is challenging in general, it is particularly difficult for low-resource languages such as Urdu. This study investigates Urdu SER in a cross-corpus setting, an area that has remained largely unexplored. We employ a cross-corpus evaluation framework across three different Urdu emotional speech datasets to test model generalization. Two standard domain-knowledge based acoustic feature sets, eGeMAPS and ComParE, are used to represent speech signals as feature vectors which are then passed to Logistic Regression and Multilayer Perceptron classifiers. Classification performance is assessed using unweighted average recall (UAR) whilst considering class-label imbalance. Results show that Self-corpus validation often overestimates performance, with UAR exceeding cross-corpus evaluation by up to 13%, underscoring that cross-corpus evaluation offers a more realistic measure of model robustness. Overall, this work emphasizes the importance of cross-corpus validation for Urdu SER and its implications contribute to advancing affective computing research for underrepresented language communities.

*Keywords Speech Emotion Recognition, SER, Affective Computing.*

## 1. Introduction

Emotions play a vital role in human communication, with speech being one of the most powerful modalities for expressing affective states [1]. Speech Emotion Recognition (SER) seeks to detect emotions from audio signals by analyzing vocal cues, which is crucial since emotions strongly influence decision-making [2].

Traditional SER approaches use domain-knowledge based handcrafted acoustic feature such as Mel Frequency Cepstral Coefficients (MFCCs) for spectral features or the extended Geneva Minimalistic Acoustic Parameter Set (eGeMAPS) and the Computational Paralinguistics Challenge (ComParE) [3, 4] feature set. Acoustic features within the two feature sets capture prosodic, spectral, and voice-quality information that are widely recognized for emotion modeling. It should be noted that whereas eGeMAPS is an optimized 88-dimensional feature set, ComParE is a large 6373-dimensional acoustic parameter set that is often considered as "brute force" dataset given the much large array of acoustic features contained within it. However, given the data driven nature of SER task, one cannot and does not know apriori which acoustic feature set will perform better for the SER classification task, where the objective to identify a categorical emotion label given an audio recording.

Despite growing research in SER, studies on underrepresented languages like Urdu spoken by over 170 million people worldwide [5], have remained limited. Existing work has focused largely on cross-lingual settings involving Urdu and western languages [6, 8], while systematic cross-corpus evaluations across multiple Urdu datasets have not been addressed to the best of our knowledge.

This study aims to fill that gap by conducting the first cross-corpus SER evaluation for Urdu using only handcrafted acoustic features. We employ two widely used feature sets, eGeMAPS and ComParE, and evaluate them with lightweight classifiers to assess their generalization capability across three Urdu emotional speech datasets. Our contributions are two-folds: this is the first systematic cross-corpus evaluation of Urdu SER using handcrafted acoustic features, where other datasets also contain Urdu language, and secondly, we propose a 3-to-1 cross-corpus validation framework to assess model generalization.

## 2. Related work

Latif et al. [6] investigated the cross-lingual performance of emotion classification for Urdu. They trained a binary SVM classifier to distinguish high and low activation/arousal classes using EMODB, SAVEE, and EMOVO datasets, and evaluated it on 400 Urdu utterances collected from TV talk shows. The cross-lingual setup achieved 71% accuracy, whereas a monolingual classifier for Urdu yielded a higher accuracy of 83.4%, highlighting the challenges of generalizing emotion recognition models across languages.

Polzehl et al. [7], investigated the cross-lingual performance of SER for German and English using recordings from German and English interactive voice response systems with five emotions. Their results showed that cross-lingual emotion recognition performs significantly worse than monolingual systems. They computed optimal acoustic and prosodic feature sets for each language and highlighted that different languages require different feature sets, indicating that a single feature selection mechanism may not generalize well across languages.

Expanding to multilingual data, Farhad et al. [8] analyzed Urdu, Arabic, and English using random forest classifiers with MFCC and pitch features, reporting a maximum recall of 79% for Urdu reflecting only moderate effectiveness of conventional acoustic features in diverse language settings.

While both Latif et al. [6] and Farhad et al. [8] highlight the challenges of Urdu SER transfer to western languages, they differ in scope and methodology: [6] focuses on direct Urdu–western cross-corpus transfer with a feature-rich SVC approach, whereas [8] adopts a broader multilingual setup with simpler acoustic features (MFCC, pitch) and a random forest classifier, yielding only moderate effectiveness. To address these challenges, Zehra et al. [9] applied ensemble learning over multilingual corpora, achieving 98.5% recall within Urdu but facing substantial drops in cross-corpus performance (63% German, 52% Italian, 39% English).

Dataset expansion has also played a significant role in advancing Urdu SER. Zaheer et al. [10] introduced SEMOUR+, a gender- and phonetically balanced emotional speech corpus containing 27,840 utterances across eight emotions (anger, boredom, disgust, neutral, surprise, happiness, sadness, and fear). They evaluated the dataset using handcrafted acoustic features such as MFCC, mel-spectrogram, and chromogram, in combination with CNN and VGG-19 classifiers. However, their results yielded relatively modest accuracies of 47% (CNN) and 52% (VGG-19).

Beyond Urdu, related work in other under-resourced languages further illustrates these challenges. For example, Mahboob et al. [11], explored SER in Sindhi using eGeMAPS and ComParE features. Reported recall scores ranged from 46.4% to 55.8%, reinforcing the difficulty of achieving robust cross-corpus generalization with handcrafted feature sets.

In Kawade and Jagtap [12], the authors proposed a cross-corpus SER framework (CCSER) that leverages multiple handcrafted acoustic features with a one-dimensional deep CNN. Using a Fire Hawk Optimization (FHO) strategy for salient feature selection, the system improved accuracy while reducing complexity. Evaluations on Hindi, Urdu, Telugu, and Kannada achieved accuracies of 58.83%, 61.75%, 69.75%, and 45.51%, respectively, showing that handcrafted features combined with selective deep models can enhance cross-lingual SER generalization.

Prior studies indicate that cross-lingual emotion recognition systems generally perform worse than monolingual systems, as emotional expression is influenced by cultural and regional factors. While some works have explored cross-corpus SER between Urdu and other languages, no prior study has focused on cross-corpus evaluation within multiple Urdu datasets. To address this gap, our methodology employs standardized preprocessing, extracts eGeMAPS and ComParE features for richer acoustic cues, and integrates self- and cross-corpus evaluation with speaker-independent folds to ensure reliable generalization

## 3. METHODOLOGY

The methodology adopted in this research is illustrated in Figure 1, where the figure provides an overview of the complete workflow, including dataset preprocessing, feature extraction, model training, and evaluation.

### 3.1. Datasets

In this study, we employed three Urdu emotional speech corpora. The first is the Latif dataset [6], a spontaneous corpus containing 400 utterances from 38 speakers, distributed across four emotions, i.e. happy, angry, sad, and neutral. The second is SEMOUR+ [10], a large, scripted and phonetically balanced dataset comprising 27,840 utterances recorded by 24 speakers across eight emotions: anger, boredom, disgust, neutral, surprise, happiness, sadness, and fear. The third is the UAM_Urdu_SER dataset, a newly collected corpus by undergraduate students at Mehran University of Engineering and Technology (credited in the acknowledgments) which is not in public domain yet, that consists of 3,953 utterances containing four emotions: happy, angry, sad, and neutral. While SEMOUR+ includes a wider set of emotion categories, the scope of our experiments is limited to these four basic emotions in order to maintain consistency across all three corpora, particularly because the UAM_Urdu_SER dataset is restricted to these emotions.

Furthermore, in this study, we simplify the classification task by adopting a binary valence-based scheme, where *Happy* and *Neutral* are grouped as positive/neutral valence (class label 0), and *Anger* and *Sad* are grouped as negative valence (class label 1). This setup allows us to focus on broad affective polarity while ensuring comparability across datasets. A summary of the datasets is provided in Table 1.

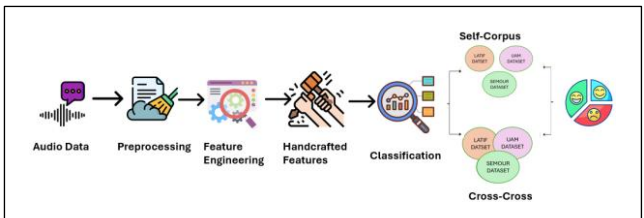

Fig. 1 Proposed Methodology

UAM_Urdu_SER annotations were performed by three raters, with majority voting used to create gold-standard (final) labels. Inter-rater agreement was determined through the Fleiss' Kappa score with a resulting score of 0.54 indicating moderate agreement [9], between the three annotators. The Kappa score highlights the challenging task of SER where human annotators may have differing views about the emotion label given the subjective nature of the task at hand

### 3.2. Preprocessing

To ensure consistency and improve the quality of audio signals, a standardized preprocessing pipeline was applied, as shown in Figure 1. Silence segments at the beginning and end of the audio files were trimmed to ensure that most of the content within the audio file is speech related to one of the emotion labels. All recordings were resampled to a uniform frequency of 16 KHz, a standard in speech processing, stereo recordings were converted to mono, and amplitude-normalized, thereby standardizing the data for subsequent feature extraction and analysis.

Table 1: Urdu Speech Emotion Datasets

| Dataset | Emotions | Samples |
|---|---|---|
| UAM_Urdu_SER Dataset (new dataset) | Happy, Anger, Sad, Neutral | 3953 |
| Latif's Dataset [6] | Happy, Anger, Sad, Neutral | 400 |
| Semour+ [10] | Happy, Anger, Sad, Neutral | 889 |

### 3.3. Feature Extraction

Following preprocessing, feature extraction was conducted using handcrafted acoustic descriptors to represent emotional cues in speech. Features were obtained using the eGeMAPS and ComParE sets from the openSMILE toolkit [13], which are widely adopted in affective computing research for their ability to capture prosodic, spectral, and voice quality characteristics. These low-level descriptors include pitch, energy, jitter, shimmer, and spectral balance, which are strongly correlated with emotional expression.

### 3.4. Data Partitioning Strategy

To ensure robust and unbiased evaluation, we adopted two complementary strategies: self-corpus evaluation and cross-corpus evaluation.

**Self-corpus evaluation:** For each dataset, we employed a Stratified Group K-Fold (SGKF) cross-validation approach. This method stratifies the folds based on emotion labels, ensuring that each split maintains balanced class distributions, while also grouping by speaker identity so that no speaker appears in both training and testing sets. Each dataset was individually partitioned into four folds. In every iteration, three folds were used for training and the remaining fold for testing, and the process was repeated until every fold had served as the test set. This design prevented speaker leakage and ensured fair comparison across models. SGKF is particularly effective in SER, where models may otherwise be overfit to speaker-specific characteristics.

**Cross-corpus evaluation:** To assess generalization across datasets, we adopted a 3-to-1 cross-corpus setup. In this scheme, one dataset (UAM_Urdu_SER, Latif, or SEMOUR+) served as the test set, while the remaining two, together with the training split of the target dataset (to ensure label consistency), were used for training. The corresponding test split of the target dataset remained unseen during training. This design allowed us to evaluate how well models trained on multiple corpora could generalize to an unseen target dataset.

### 3.5. Classification

After feature extraction, the handcrafted acoustic features were used as inputs to two classifiers: Logistic Regression (LR), and Multilayer Perceptron (MLP). Hyperparameters of the models were optimized using grid/randomized search, and performance was evaluated through cross-validation to ensure robust results.

## 4. RESULT & ANALYSIS

This section presents the performance of Urdu SER models under Self-corpus and Cross-corpus conditions.

### 4.1. Evaluation Metric

To evaluate the performance of the classification models, mean recall, often called unweighted average recall (UAR), was computed. Considering the class imbalance present in the datasets, mean recall was adopted as the primary evaluation metric, as it better reflects the model's ability to correctly identify minority emotion classes.

### 4.2 Self-Corpus Evaluation

In this set of experiments, models were trained and tested on the same dataset to evaluate their performance under a Self-corpus setting.

The performance of handcrafted acoustic features was first evaluated using Logistic Regression (LR) under Self-corpus settings, as presented in Table 2 and a visual comparison is provided in Figure 2 and Figure 3, which show that the SEMOUR dataset achieved the highest recall mean, with eGeMAPS reaching 81.53% and ComParE 78.29%, indicating that handcrafted features generalize well when sufficient data diversity is present. For the UAM_Urdu_SER dataset, performance was moderate, with ComParE slightly outperforming eGeMAPS (64.64% vs. 61.91%). In contrast, for the Latif dataset, eGeMAPS achieved better results 64.84% compared to ComParE 56.35%, highlighting the effectiveness of prosodic and spectral descriptors in smaller datasets.

Table 2: Logistic Regression Performance reported in terms of UAR (%).

| Dataset | eGeMAPS | | ComParE | |
|---|---|---|---|---|
| | Self-Corpus | Cross-Corpus | Self-Corpus | Cross-Corpus |
| Latif | 64.84 | 51.82 | 56.35 | 57.57 |
| SEMOUR | 81.53 | 71.84 | 78.29 | 73.26 |
| UAM | 61.91 | 59.72 | 64.64 | 63.03 |

When using the Multi-Layer Perceptron (MLP) classifier, a consistent trend was observed across the three Urdu datasets, as shown in Table 3. The SEMOUR dataset again yielded the highest performance, with 79.13% recall mean for eGeMAPS and 77.98% for ComParE, confirming its robustness for handcrafted feature-based SER. For the UAM_Urdu_SER dataset, the results were again moderate, with ComParE slightly outperforming eGeMAPS (63.83% vs. 61.87%). On the smaller Latif dataset, eGeMAPS performed notably better 60.40% compared to ComParE 55.36%, highlighting its suitability for datasets with limited size and balanced samples.

Table 3: Multi-Layer Perceptron Performance reported in terms of UAR (%).

| Dataset | eGeMAPS | | ComParE | |
|---|---|---|---|---|
| | Self-Corpus | Cross-Corpus | Self-Corpus | Cross-Corpus |
| Latif | 60.40 | 59.84 | 55.36 | 55.09 |
| SEMOUR | 79.13 | 78.06 | 77.98 | 74.76 |
| UAM | 61.87 | 61.11 | 63.83 | 62.67 |

### 4.3 Cross-Corpus Evaluation

Cross-corpus experiments were conducted to assess the generalizability of handcrafted acoustic features across unseen Urdu datasets. Unlike Self-corpus evaluation, where training and testing are performed on the same dataset, cross-corpus evaluation trains on one dataset and tests on another, providing a more realistic measure of robustness. We employed a 3-to-1 cross-corpus setting using three distinct Urdu emotional speech datasets: UAM_Urdu_SER, Latif, and SEMOUR. In each experimental iteration, one dataset was designated as the target corpus for testing, while the other two datasets, along with the training portion of the target corpus itself, were used for training.

For the Logistic Regression (LR) classifier, the SEMOUR dataset achieved the highest recall values, with ComParE 73.26% slightly outperforming eGeMAPS 71.84%, as presented in Table 2. The UAM_Urdu_SER dataset also showed competitive performance, where ComParE reached 63.03% compared to 59.72% for eGeMAPS. In contrast, the Latif dataset yielded comparatively lower results, with eGeMAPS at 51.82% and ComParE at 57.57%, indicating challenges in cross-corpus generalization on smaller datasets.

Similarly, for the Multi-Layer Perceptron (MLP) classifier, shown in Table 3, the SEMOUR dataset again achieved the highest recall, with eGeMAPS at 78.06% and ComParE at 74.76%, demonstrating its robustness across corpora. The UAM_Urdu_SER dataset produced balanced results, where ComParE 62.67%. slightly outperformed eGeMAPS 61.11%. On the Latif dataset, both feature sets struggled, with eGeMAPS achieving 59.84% and ComParE 55.09%, reaffirming the difficulty of cross-corpus generalization under limited sample size conditions.

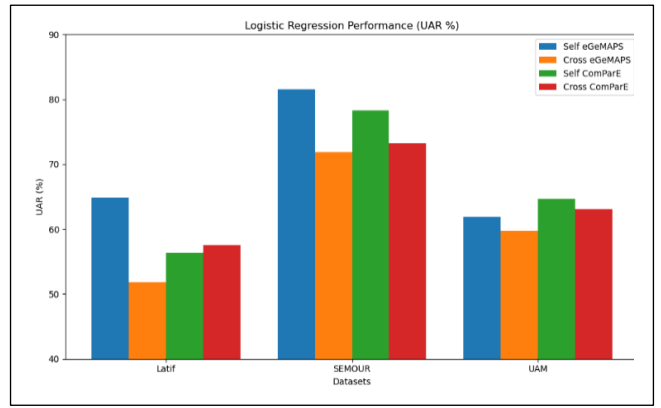

Fig. 2 Logistic Regression UAR (%) across datasets using eGeMAPS and ComParE.

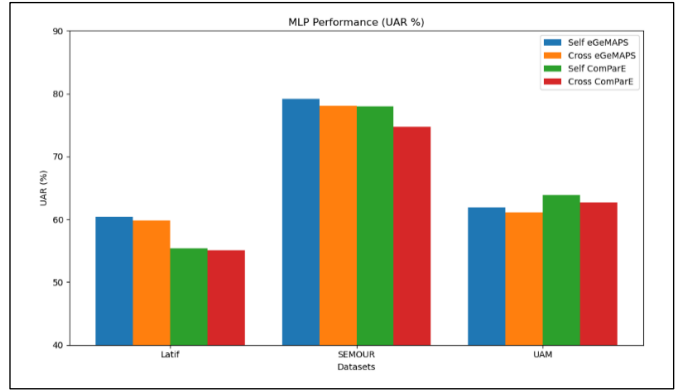

Fig. 3 MLP UAR (%) across datasets using eGeMAPS and ComParE.

### 5. DISCUSSION

This study evaluated handcrafted acoustic features (eGeMAPS and ComParE) for Urdu SER across three datasets: UAM_Urdu_SER, LATIF, and SEMOUR, under both Self-corpus and cross-corpus conditions. Self-corpus evaluations demonstrated consistent performance across cross-validation folds, indicating low variability of the classifiers when trained and tested on the same dataset. Among the three, SEMOUR achieved the highest recall values (above 78%), reflecting the advantages of larger and more balanced datasets. UAM_Urdu_SER also showed competitive results (around 61–64%), whereas the smaller LATIF dataset achieved lower recall (55–65%), highlighting the impact of dataset size and diversity on classifier performance.

In contrast, cross-corpus evaluations revealed a clear performance drop across all datasets due to domain shift, underscoring the gap between within-dataset learning and real-world generalization. SEMOUR again demonstrated the strongest transferability, while LATIF exhibited the weakest, suggesting that limited data and recording variability restrict the robustness of handcrafted features. Between the two feature sets, no clear dominance was observed: eGeMAPS sometimes outperformed ComParE (e.g., LATIF with MLP), while ComParE showed slight advantages in other cases (e.g., UAM with LR), indicating that feature–classifier compatibility is context-dependent.

Overall, the results reveal that while handcrafted features capture meaningful acoustic cues for emotion recognition, their generalization across corpora remains limited. Dataset size, balance, and recording conditions significantly influence performance, and the observed performance drop in cross-corpus experiments underscores the challenges of developing robust Urdu SER systems using handcrafted approaches.

## 6. Conclusions, Limitations and Future Work

This study marks a significant step forward in Urdu SER by introducing the first comprehensive cross-corpus evaluation of handcrafted acoustic features, providing valuable insights into their strengths and limitations in low-resource settings. Using well-established feature sets such as eGeMAPS and ComParE, the experiments demonstrated that while handcrafted features can achieve competitive performance in Self-corpus evaluations, their generalization across unseen datasets remains a major challenge. This performance gap between Self- and cross-corpus results highlights the strong influence of dataset-specific factors such as recording conditions, speaker variability, and annotation inconsistencies.

The hypothesis was that positive and negative valence categories generalize well across corpora but that was not supported, further emphasizing the limited shared emotional structure among existing Urdu datasets. This finding underscores the importance of dataset standardization and the need to establish unified labeling schemes for future SER research in Urdu.

This work is limited by the scarcity and inconsistency of available Urdu emotional speech datasets, which vary in size, speaker diversity, and balance of emotion categories. Moreover, the experiments were restricted to only four emotion classes (happy, angry, sad, and neutral), whereas in reality human emotions are more diverse and often overlapping. These limitations restrict the robustness of handcrafted features, particularly in cross-corpus evaluation where domain shift significantly reduces performance.

Future research should prioritize the development of larger, balanced, and consistently annotated Urdu emotion datasets to support more reliable evaluation. In addition, adopting multi-label classification frameworks would allow systems to capture the overlapping and co-occurring nature of human emotions, providing a more realistic representation of affective states. Another promising direction is the integration of handcrafted features with deep acoustic embeddings, enabling complementary representations that may enhance cross-corpus generalization and bridge the performance gap between traditional and modern SER approaches.

## Acknowledgements
We gratefully acknowledge Azeem Memon and Mashood Mangrio for their contribution in the development of the UAM_Urdu_SER dataset, which was collected as part of our undergraduate work at Mehran University of Engineering and Technology (MUET), Pakistan. We also extend our appreciation to Hafsa Sadaqat for her valuable contribution to the dataset annotation process.